\documentclass[twocolumn,floats,floatfix,superscriptaddress,aps,pra]{revtex4}
\usepackage{amsmath}
\usepackage{graphicx}
\usepackage{multirow}
\usepackage{booktabs}
\usepackage{afterpage}

\usepackage{array}
\newcolumntype{L}[1]{>{\raggedright\let\newline\\\arraybackslash\hspace{0pt}}m{#1}}
\newcolumntype{C}[1]{>{\centering\let\newline\\\arraybackslash\hspace{0pt}}m{#1}}
\newcolumntype{R}[1]{>{\raggedleft\let\newline\\\arraybackslash\hspace{0pt}}m{#1}}

\def\be{ \begin{equation} }
\def\beal{ \begin{align} }
\def\ee{ \end{equation} }
\def\bea{ \begin{eqnarray} }
\def\eea{ \end{eqnarray} }
\def\bse{ \begin{subequations} }
\def\ese{ \end{subequations} }
\def\ba{ \begin{array} }
\def\ea{ \end{array} }
\def\bwt{ \begin{widetext} }
\def\ewt{ \end{widetext} }

\def\U{\mathbf{U}}
\def\F{\mathbf{F}}

\def\i{{\rm{i}}}

\def\e{{\rm{e}}}

\def\ket#1{\vert #1\rangle}

\def\matrix22#1#2#3#4{\left[ \begin{array}{cc} #1 & #2 \\ #3 & #4 \end{array}\right]}

\def\bt{\begin{tabular}}
\def\et{\end{tabular}}

\usepackage{mathptmx}
\usepackage{psfrag,graphicx}
\usepackage{dcolumn}
\usepackage{amsmath,amssymb,amsfonts}
\usepackage{bm}
\usepackage{color}
\usepackage{latexsym}
\usepackage{epstopdf}
\usepackage{color}
\usepackage[english]{babel}
\usepackage{natbib}
\usepackage{appendix}
\usepackage[normalem]{ulem}

\usepackage{xy}
\xyoption{matrix}
\xyoption{frame}
\xyoption{arrow}
\xyoption{arc}
\usepackage{ifpdf}
\ifpdf
\else
\PackageWarningNoLine{Qcircuit}{Qcircuit is in PS mode. The Xy-pic options ps and dvips are loaded. If you wish to use other PS drivers for Xy-pic, you must modify the code in Qcircuit.tex}
\xyoption{ps}
\xyoption{dvips}
\fi

\entrymodifiers={!C\entrybox}
\newcommand{\qw}[1][-1]{\ar @{-} [0,#1]}
\newcommand{\qwx}[1][-1]{\ar @{-} [#1,0]}

\newcommand{\measure}[1]{*+[F-:<.9em>]{#1} \qw}

\newcommand{\control}{*!<0em,.025em>-=-<.2em>{\bullet}}

\newcommand{\ctrl}[1]{\control \qwx[#1] \qw}

\newcommand{\targ}{*+<.02em,.02em>{\xy ="i","i"-<.39em,0em>;"i"+<.39em,0em> **\dir{-}, "i"-<0em,.39em>;"i"+<0em,.39em> **\dir{-},"i"*\xycircle<.4em>{} \endxy} \qw}

\newcommand{\multigate}[2]{*+<1em,.9em>{\hphantom{#2}} \POS [0,0]="i",[0,0].[#1,0]="e",!C *{#2},"e"+UR;"e"+UL **\dir{-};"e"+DL **\dir{-};"e"+DR **\dir{-};"e"+UR **\dir{-},"i" \qw}
\newcommand{\ghost}[1]{*+<1em,.9em>{\hphantom{#1}} \qw}

\newcommand{\rstick}[1]{*!L!<-.5em,0em>=<0em>{#1}}
\newcommand{\lstick}[1]{*!R!<.5em,0em>=<0em>{#1}}

\newcommand{\Qcircuit}{\xymatrix @*=<0em>}

\begin{document}

\author{Boyan T. Torosov}
\affiliation{Institute of Solid State Physics, Bulgarian Academy of Sciences, 72 Tsarigradsko chauss\'{e}e, 1784 Sofia, Bulgaria}
\author{Nikolay V. Vitanov}
\affiliation{Department of Physics, St Kliment Ohridski University of Sofia, 5 James Bourchier Blvd, 1164 Sofia, Bulgaria}

\title{Narrowband composite two-qubit phase gates}

\date{\today}

\begin{abstract}

We propose a method to construct composite two-qubit gates with narrowband profiles with respect to the spin-spin coupling. 
The composite sequences are selective to the variations in the amplitude and duration of the spin-spin coupling, and can be used for highly-selective qubit addressing with greatly reduced cross talk, quantum logic spectroscopy, and quantum sensing. 

\end{abstract}

\maketitle

%\section{Introduction}\label{sec-introduction}

Composite pulses are a powerful quantum control technique, which offers a large variety of excitation profile features, such as high fidelity, robustness, sensitivity, etc. 
A composite pulse is actually a sequence of pulses with well-defined and different relative phases, used as control parameters to achieve a certain objective.
First developed in nuclear magnetic resonance (NMR) \cite{NMR} and its analogue even earlier in polarization optics \cite{PolarizationOptics}, their efficiency was soon largely acknowledged.
In the last two decades, they have been successfully applied in other areas, such as trapped ions \cite{Gulde2003, Mount2015, Schmidt-Kaler2003, Haffner2008, Timoney2008, Monz2009, Zarantonello2019, Shappert2013}, neutral atoms \cite{Rakreungdet2009, Demeter2016}, quantum dots \cite{Wang2012, Eng2015, Kestner2013, Wang2014, Zhang2017, Hickman2013}, NV centers in diamond \cite{Rong2015}, doped solids \cite{Schraft2013, Genov2014, Genov2017, Bruns2018}, superconducting qubits \cite{SteffenMartinisChuang, TorosovIBM}, optical clocks \cite{Zanon-Willette2018}, atom optics \cite{Butts2013, Dunning2014, Berg2015}, magnetometry \cite{Aiello2013}, optomechanics \cite{Ventura2019}, etc.

Traditionally, composite pulses are primarily used to achieve \textit{broadband} excitation profiles \cite{Wimperis1994, Torosov2011, Torosov2015, TorosovIBM} with high fidelity, robust to deviations in certain experimental parameters, such as the amplitude, frequency and duration of the external driving pulsed fields. 
Recently, composite pulses which are robust to deviations in \textit{any} parameter have been designed and demonstrated \cite{Genov2014}.
However, one may also use composite pulses to achieve \textit{narrowband} profiles \cite{Vitanov2011, Torosov2015, Wimperis1994, TorosovIBM} for highly selective excitation only within a narrow range of a certain parameter value. 
A combination of broadband and narrowband features are offered by \textit{passband} composite pulses \cite{Wimperis1994, Kyoseva2013}, which lie between these two extremes.

The narrowband excitation profiles --- an often overseen unique feature of composite pulses that no other quantum control technique offers --- finds interesting applications, e.g., in spatial localization of excitation for higher-resolution imaging, as implemented already in the NMR age of composite pulses \cite{NB-NMR, Wimperis1994}.
In polarization optics, the equivalent of narrowband pulses are used for polarization filters \cite{Dimova2014}, where frequency resolution can be pushed to 1 nm or even better.
NB pulses can be a very promising tool in selective spatial addressing of tightly spaced trapped ions in 1D or 2D ion crystals or atoms in optical lattices by tightly focused laser beams \cite{Ivanov2011}.
In this manner, NB pulses can significantly reduce the unwanted cross talk to neighbouring qubits in a quantum register while implementing quantum gates on the desired qubit(s).
In this manner, because unwanted cross talk is one of the main contributor to the error budget, narrowband composite pulses can improve the fidelity of the quantum circuit. 
They can be used also in selective addressing of a particular vibrational sideband frequency mode of a trapped ion, thereby making it possible to precisely determine an arbitrary phonon (thermal or non-thermal) distribution of a trapped ion. %% \cite{Stockholm}. 

Narrowband pulses have been designed for complete (X gate) \cite{Wimperis1994,Ivanov2011, Vitanov2011, Torosov2015} and partial (Hadamard gate and general rotations) \cite{Wimperis1994, Ivanov2011, Torosov2020} but narrowband versions of a two-qubit gate %controlled-phase or controlled-NOT gates
have not been developed hitherto.
The two-qubit gates are an essential part of any quantum computing protocol. 
In combination with a set of three single-qubit gates, they can form a universal set of gates, capable of producing any unitary operation, corresponding to a quantum algorithm, over a register of arbitraily many qubits. 
Some popular two-qubit gates are the CNOT, CPHASE/CZ, and fSim/iSWAP gates. 
The CPHASE and CNOT gates are locally equivalent, while the CNOT can be produced by a pair of fSim gates plus a few single-qubit gates \cite{google-supremacy}. 

Methods to generate broadband and passband two-qubit gates by using composite sequences have been previously explored in the literature \cite{CompositeTwoQubitGates}.
In this paper, we introduce a method to derive highly-sensitive narrowband composite controlled-phase gates by using sequences of  two-qubit $\sigma_x\sigma_x$ interactions, interleaved with single-qubit phase gates with appropriately chosen phases. 
We derive such sequences of up to $N=11$ pulses and perform simulations to test the profile of the fidelity as a function of the deviation in the interaction strength or duration.
In this manner, by designing the hitherto unavailable narrowband two-qubit quantum gate, we complete the library of available composite gates (single-qubit and two-qubit quantum gates, each with broadband, narrowband and passband fidelity profiles), thereby supplying the experimenter with a variety of all possible choices suitable for a particular experimental situation.
%We discuss possible applications of the method in the conclusion section.

%%%%%%%%%%%%%%%%%%%%%%%%%%%%%%%%%%%%%%%%%%%%%%%%%%%%%%%
%\section{Two-qubit control phase gate}\label{sec-two-qubit}
\textbf{Derivation method.} In this section, our goal is to produce a gate of the type
\be\label{Utarget}
U(\theta) = \e^{\i\theta\sigma_x\sigma_x},
\ee
with a narrowband fidelity profile, by using a sequence of (ordinary, non-narrowband) XX propagators of the type \eqref{Utarget}.
Here $\theta=JT$ is a rotation angle generated by a coupling strength $J$ acting for an interaction duration of $T$.

We note here that traditionally a CPHASE gate is defined as $\text{diag}[1,1,1,\e^{\i\varphi}]$, which is locally equivalent to the $\exp(\i\theta\sigma_z\sigma_z)$ gate, with $\varphi=4\theta$. 
The original CPHASE gate can be obtained from  $U(\theta)$ of Eq.~\eqref{Utarget} as 
\be
\e^{\i\varphi/4}\e^{-\i(\varphi/4)\sigma_{z,1}}\e^{-\i(\varphi/4)\sigma_{z,2}}\e^{\i(\varphi/4)\sigma_{z,1}\sigma_{z,2}},
\ee
and an Hadamard transformation between the $Z$ and $X$ bases, where $\sigma_{z,k}$ is the Pauli-Z operator, applied on qubit $k$.
For mathematical convenience, we are going to work in the rotated $X$ basis.
%, where the $X$ and $Z$ bases are connected by an Hadamard transformation.

In our derivation, we are going to assume that the rotation angle $\theta$ has some relative deviation $\epsilon$ from the perfect (nominal) value $\Theta$, and therefore $\theta=\Theta(1+\epsilon)$. 
Furthermore, we are going to assume two different values for the target angle, namely $\Theta=\pi/4$, which is usually used to create a CZ (and CNOT) gate, and $\Theta=\pi/2$.
A composite sequence of length $N$ produces the propagator
\be\label{UN}
\U^{(N)}(\theta) = \U_{\phi_N}(\theta_N)\cdots \U_{\phi_2}(\theta_2) \U_{\phi_1}(\theta_1), 
\ee
where
\be\label{Uphi}
\U_{\phi}(\theta) = \F(-\phi) \U(\theta) \F(\phi)
\ee
is a phase-shifted propagator, and the single-qubit phase gate 
\be\label{Phase_gate}
\F(\phi) = \exp(\i\phi \sigma_z)
\ee 
is applied on only one of the qubits. 
To be specific, we are going to apply these phase shifts on qubit 1.
Because two adjacent phase gates \eqref{Phase_gate} produce a phase gate with the sum of their phases, we can write Eq.~\eqref{UN} as [cf. Eq.~\eqref{Uphi}]
\begin{align}
\U^{(N)}(\theta) &= \F(-\phi_N) \U(\theta_N) \F(\phi_N-\phi_{N-1}) \U(\theta_N) \cdots \notag\\
&\cdots \F(\phi_3-\phi_2) \U(\theta_2) \F(\phi_2-\phi_1) \U(\theta_1) \F(\phi_1). 
\end{align}
These can be schematically depicted as
\begin{widetext}
\be
\Qcircuit @C=1em @R=.7em {
& \measure{\phi_1} & \multigate{1}{\theta_1} & \measure{\phi_2-\phi_1} & \multigate{1}{\theta_2} & \measure{\phi_3-\phi_2} & \qw & & \qw & & \qw & & \measure{\phi_{N-1}-\phi_{N-2}} & \multigate{1}{\theta_{N-1}} & \measure{\phi_{N}-\phi_{N-1}} & \multigate{1}{\theta_N} & \measure{-\phi_{N}} &
\\
& \qw & \ghost{\theta_1} & \qw & \ghost{\theta_2} & \qw & \qw & & \qw & & \qw & & \qw & \ghost{\theta_{N-1}} & \qw & \ghost{\theta_N} & \qw &
}
\ee
\end{widetext}
where a rounded box means the single-qubit phase gate \eqref{Phase_gate} and a square box is the two-qubit XX gate \eqref{Utarget}.
The first and the last phase gates are not essential, but we keep them for the sake of symmetry.
%do not change the fidelity of the gate but they are needed to produce the target gate \eqref{Utarget}.

The narrowband CPs, derived by our approach have the following property: they produce the target gate $\U(\Theta)$ at $\epsilon=0$, while producing the identity $\mathbf{1}$ for $\epsilon=\pm 1$, and some vicinities of these points, up to a certain order of accuracy. 
These properties are achieved by imposing the conditions
\begin{subequations}\label{NB-eqs}
\begin{align}
\left.[U^{(N)}(\theta)-U(\theta)]\right|_{\epsilon=0}=0& \\
\frac{\partial^l}{\partial\epsilon^l}\left.[U^{(N)}(\theta)-\mathbf{1}]\right|_{\epsilon=\pm 1}=0&  \quad
(l=0,1,\ldots,n),
\end{align}
\end{subequations}
with $n$ being the order of compression of the narrowband pulses. 
The first of these equations guarantees that for zero deviation ($\epsilon=0$), the composite gate will be the XX gate of Eq.~\eqref{Utarget} with the nominal angle $\Theta$.
The second equation, which is actually a set of $n$ equations, defines the narrowband nature of the fidelity profile --- the higher the value of $n$, the greater the squeezing of the profile, and the stronger the narrowband effect.

The performed numerical simulations show that our sequences must be of the form of a $\pi/4$ gate, followed by a sequence of an \textit{odd} number of $\pi/2$ gates, and in the end another $\pi/4$ gate,
\be
U^{(N)}(\theta) = \U_{\phi_N}\left(\tfrac{\pi}{4}\right) \U_{\phi_{N-1}}\left(\tfrac{\pi}{2}\right) \cdots  \U_{\phi_2}\left(\tfrac{\pi}{2}\right) \U_{\phi_1}\left(\tfrac{\pi}{4}\right).
\ee
By solving Eqs.~\eqref{NB-eqs} we derive the composite phases $\phi_k$, forming our narrowband composite gates. 

In order to test the performance of our composite sequences, we define the fidelity of the composite gate as
\be\label{fidelity}
F = \tfrac14 \text{Tr}\left[ U^{\dagger}(\theta) U^{(N)}(\theta) \right].
\ee

We proceed below by first explicitly examining the simplest case of composite sequences of $N=3$ components, then $N=5$, and then the general case.

\textbf{Three-component composite sequences.}
For sequences of $N=3$ gate segments $\U_{\phi}(\theta)$ of Eq.~\eqref{Uphi} and $\Theta=\pi/2$, Eqs.~\eqref{NB-eqs} generate, up to the first order ($n=1$) in $\varepsilon$, the set of equations
\bse
\begin{align}
&\sin(\phi_1-\phi_3) = 0, \\
&e^{4 i \phi_1 - 2 i \phi_2} - e^{2 i \phi_2} + 2 = 0,\\
&e^{4 i \phi_1} + e^{4 i \phi_2} = 0,
\end{align}
\ese
and a set of complex conjugated of these equations, which are omitted due to redundancy.
These equations are readily solved, and they have multiple solutions,
\bse
\begin{align}
\phi_1 &= \frac{(2k_1+1)\pi}{4}, \quad (k_1=0,1,2,3),\\
\phi_2 &= k_2 \pi, \quad (k_2=0,1),\\
\phi_3 &= \phi_1 + k_3 \pi, \quad (k_3=0,1).
\end{align}
\ese
Similar results can be obtained for $\Theta=\pi/4$.

All of these solutions produce the same fidelity profile.
Unfortunately, three-component composite sequences do not offer any advantage over the single XX gate \eqref{Utarget}, because they produce identical fidelity profiles, as the XX gate \eqref{Utarget} also suppresses first-order ($n=1$) deviations in $\theta$.

%FIGURE
\begin{figure}[tbph]
	\includegraphics[width=0.8\columnwidth]{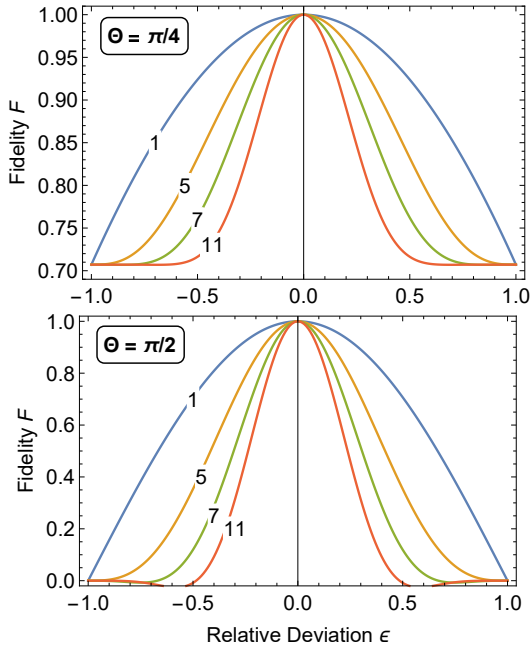}
	\caption{Fidelity of composite CPHASE gate as a function of deviation $\epsilon$ of the target angle $\Theta$ for sequences of $N=1,5,7,11$ pulses.  The target angle is $\Theta=\pi/4$ (top) and $\Theta=\pi/2$ (bottom).}
	\label{Fig:Fidelity}
\end{figure}
%FIGURE

\textbf{Five-component composite sequences.}
For sequences of $N=5$ gate segments $\U_{\phi}(\theta)$ of Eq.~\eqref{Uphi} and target angle $\Theta=\pi/4$, Eqs.~\eqref{NB-eqs} generate, up to the second order in $\varepsilon$, the set of equations
\bse
\begin{align}
& e^{4i (\phi_2+\phi_4)} - e^{2i (\phi_1+2\phi_3+\phi_5)} + \sqrt{2}\, e^{2i (\phi_2+\phi_3+\phi_4)} = 0,\\
& e^{4i (\phi_2+\phi_4)} + e^{2i (\phi_1+2\phi_3+\phi_5)} - \sqrt{2}\, e^{2i (\phi_1+\phi_2+\phi_3+\phi_4)} = 0,\\
& \sin (\phi_1-\phi_5) = 0,\\
& e^{2i \phi_1} + 2 e^{2 i \phi_2}+2 e^{2 i \phi_3}+2 e^{2 i \phi_4}+e^{2 i \phi_5} = 0.
\end{align}
\ese
These equations have four solutions,
\begin{subequations}
\begin{align}
\{\phi_1,\phi_2,\phi_3,\phi_4,\phi_5\} &= \{\tfrac14,\tfrac{5}{16},\tfrac34,\tfrac{13}{16},\tfrac14\}\pi \label{CPquarter-5a},\\
\{\phi_1,\phi_2,\phi_3,\phi_4,\phi_5\} &= \{\tfrac14,\tfrac{13}{16},\tfrac34,\tfrac{5}{16},\tfrac14\}\pi ,\\
\{\phi_1,\phi_2,\phi_3,\phi_4,\phi_5\} &= \{\tfrac14,\tfrac{13}{16},\tfrac34,\tfrac{21}{16},\tfrac14\}\pi ,\\
\{\phi_1,\phi_2,\phi_3,\phi_4,\phi_5\} &= \{\tfrac14,\tfrac{21}{16},\tfrac34,\tfrac{13}{16},\tfrac14\}\pi .
\end{align}
\end{subequations}
The second and fourth solutions are the inverted first and second solutions, respectively; all solutions produce the same fidelity profile.

For $N=5$ and $\Theta=\pi/2$, Eqs.~\eqref{NB-eqs} generate the set of equations
\bse
\begin{align}
%& e^{2i (\phi_2-\phi_3+\phi_4)} - e^{2i (\phi_1-\phi_2+\phi_3-\phi_4+\phi_5)} + 2 = 0,\\
%& e^{2i (\phi_1-2\phi_2+2\phi_3-2\phi_4+\phi_5)} + 1 = 0,\\
& e^{4i (\phi_2+\phi_4)} - e^{2i (\phi_1+2\phi_3+\phi_5)} + 2 e^{2i (\phi_2+\phi_3+\phi_4)} = 0,\\
& e^{4i (\phi_2+\phi_4)} + e^{2i (\phi_1+2\phi_3+\phi_5)} = 0,\\
&\sin (\phi_1-\phi_5) = 0,\\
&e^{2 i \phi_1}+2 e^{2 i \phi_2}+2 e^{2 i \phi_3}+2 e^{2 i \phi_4}+e^{2 i \phi_5} = 0.
\end{align}
\ese
These equations have two solutions,
\begin{subequations}
\begin{align}
\{\phi_1,\phi_2,\phi_3,\phi_4,\phi_5\} &= \{\tfrac14,\tfrac38,\tfrac34,\tfrac78,\tfrac14\}\pi , \label{CPhalf-5a}\\
\{\phi_1,\phi_2,\phi_3,\phi_4,\phi_5\} &= \{\tfrac14,\tfrac78,\tfrac34,\tfrac38,\tfrac14\}\pi .
\end{align}
\end{subequations}
The second solution is the inverted first solution and it produces the same fidelity profile.

The first solutions of the above sets \eqref{CPquarter-5a} and \eqref{CPhalf-5a} are implemented with the following quantum circuits, %\begin{subequations}
%\begin{align}
%\phi_1=\phi_5=-\phi_3=\pi/4 , \\
%\phi_{2} = 5\pi/16,\quad \phi_4 = -3\pi/16 ,
%\end{align}
%\end{subequations}
%for $\Theta=\pi/4$ and
\begin{widetext}
\bse
\begin{align}
&
\Qcircuit @C=1em @R=.7em {
& \measure{\frac{\pi}{4}} & \multigate{1}{\frac{\pi}{4}} & \measure{\frac{\pi}{16}} & \multigate{1}{\frac{\pi}{2}} & \measure{\frac{7\pi}{16}} & \multigate{1}{\frac{\pi}{2}} & \measure{\frac{\pi}{16}} & \multigate{1}{\frac{\pi}{2}} & \measure{-\frac{9\pi}{16}} & \multigate{1}{\frac{\pi}{4}} & \measure{-\frac{\pi}{4}} & \qw
 \\
& \qw & 
\ghost{\frac{\pi}{4}} & \qw & \ghost{\frac{\pi}{2}} & \qw & \ghost{\frac{\pi}{2}} & \qw & \ghost{\frac{\pi}{2}} & \qw & \ghost{\frac{\pi}{4}} & \qw & \qw
}
\\ \notag \\
&
\Qcircuit @C=1em @R=.7em {
& \measure{\frac{\pi}{4}} & \multigate{1}{\frac{\pi}{4}} & \measure{\frac{\pi}{8}} & \multigate{1}{\frac{\pi}{2}} & \measure{\frac{3\pi}{8}} & \multigate{1}{\frac{\pi}{2}} & \measure{\frac{\pi}{8}} & \multigate{1}{\frac{\pi}{2}} & \measure{-\frac{5\pi}{8}} & \multigate{1}{\frac{\pi}{4}} & \measure{-\frac{\pi}{4}} & \qw
\\
& \qw & 
\ghost{\frac{\pi}{4}} & \qw & \ghost{\frac{\pi}{2}} & \qw & \ghost{\frac{\pi}{2}} & \qw & \ghost{\frac{\pi}{2}} & \qw & \ghost{\frac{\pi}{4}} & \qw & \qw
}
\end{align}
\ese
\end{widetext}
where, as before, a rounded box with a number $\phi_k$ means the single-qubit phase gate \eqref{Phase_gate} with the phase $\phi_k$, and a square box means the two-qubit XX gate \eqref{Utarget} with the value of $\Theta$ inside.

Figure \ref{Fig:Fidelity} shows the fidelity of Eq.~\eqref{fidelity} plotted versus the deviation $\varepsilon$ in the target angle $\Theta$ for a single XX gate \eqref{Utarget} and a few composite XX gates for both target angles $\Theta = \pi/4$ (top) and $\Theta = \pi/2$ (bottom).
Clearly, the five-component composite sequences shrinks the fidelity profile by about $\frac13$, thereby enhancing the selectivity of the gate.

%######################################################
\begin{table}[tb]
\caption{Phases of composite sequences for narrowband two-qubit gates, in units of $\pi$. 
The target angle $\Theta$ is $\pi/4$ (top) and $\pi/2$ (bottom)} % title of Table
\begin{tabular}{ll} % centered columns (4 columns)
\hline %inserts double horizontal lines
$N$ & Phases $\phi_1,\ldots,\phi_N$ \\ %  inserts table
\hline % inserts single horizontal line
%$\pi$ & 0 \\
 5 & $0.25, 0.3125, 0.75, 0.8125, 0.25$ \\
7 & $-0.75, -0.5006, 0.6743, 1.2249, -0.0244, -0.1994, -0.75$ \\
 9 & $0.25, -1.1584, 0.4493, 1.4203, 0.8155,$\\ 
  & $0.3416, -1.0507, -0.0797, 0.25$ \\
11 & $0.25, 0.7984, -0.1473, 0.4942, 0.1257, 1.2468,$ \\ 
   & $0.6985, 0.6441, -0.9973, 0.3711, 0.25$ \\
\hline %inserts single line
%$\pi$ & 0 \\
 5 & $0.25, 0.375, 0.75, 0.875, 0.25$ \\
7 & $0.25, -0.0475, -0.3857, -0.6763, -0.3789, -0.0407, 0.25$ \\
 9 & $0.25, 0.9480, -0.8148, -1.3878, 0.7500,$\\ 
  & $0.4480, 0.6852, -0.8878, 0.25$ \\
11 & $0.25, 0.8690, 0.2018, 0.4679, -0.3659, -0.0021,$ \\ 
  & $-0.6212, 0.0461, -0.2200, 0.6138, 0.25$ \\
\hline %inserts single line
\end{tabular}
\label{table-phases} % is used to refer this table in the text
\end{table}
%######################################################

\textbf{Longer composite sequences.}
For longer sequences, the first and the last phase remain equal to $\pi/4$, but it does not appear possible to find analytic expressions for the other phases of the single-qubit phase gates.
However, the respective equations can be solved numerically.
As for $N=5$ segments, there are multiple solutions producing the same fidelity profiles.
The values of the phases for some representative composite sequences are given in Table~\ref{table-phases}.

In Fig.~\ref{Fig:Fidelity} we plot the fidelity of the composite sequences of up to 11 components, producing a narrowband CPHASE gate, as a function of the deviation parameter $\epsilon$. 
We notice that we have a CPHASE gate for $\epsilon=0$, while a robust identity is produced in the vicinities of $\epsilon=\pm 1$.
As expected, as the length $N$ of the composite sequence increases, the fidelity profiles shrink ever more. 
In particular, for $N=11$, the full-width-at-half-maximum of the fidelity profile is about a factor of 3 more narrow than for a single gate.
Moreover, the cross talk is essentially eliminated for $\epsilon \in [-1,-0.5]$, meaning that even if a qubit ``sees'' up to 50\% of the nominal value of the coupling it will remain unchanged.

\textbf{Conclusion.}
In conclusion, we presented composite two-qubit gates with narrowband fidelity profiles. 
Such composite sequences produce the desired gate when perfect values for the interaction strength and duration are selected, but leave the systems unaffected in the wings of the fidelity profile. 
Therefore, these gates can be used for quantum information processing in quantum systems where high selectivity is needed, e.g., to suppress unwanted cross talk in laser-addressed trapped ions and atoms.
Due to their enhanced sensitivity, they can also be useful in quantum logic spectroscopy and quantum sensing.

This work is supported by the European Commission's Horizon-2020 Flagship on Quantum Technologies project 820314 (MicroQC).

%%%%%%%%%%%%%%%%%%%%%%%%%%%%%%%%%%%%%%%%%%%%%%%%%%%%%%%%%%%%%%%%%%%%%%%%%%%
%%%%%%%%%%%%%%%%%%%%%%%%%%%%%%%%%%%%%%%%%%%%%%%%%%%%%%%%%%%%%%%%%%%%%%%%%%%
%%%%%%%%%%%%%%%%%%%%%%%%%%%%%%%%%%%%%%%%%%%%%%%%%%%%%%%%%%%%%%%%%%%%%%%%%%%

\end{document}